\documentclass[12pt]{iopart}

\usepackage{iopams}
\usepackage{graphicx}
\begin{document}

\maketitle{\bf Short-range nuclear effects on axion emissivities by nucleon-nucleon bremsstrahlung}

\author{S. Stoica$^{a,b}$, B. Pastrav$^b$ and A.C. Scafes$^a$}
\address{$^a$ Horia~ Hulubei~ National~ Institute~ of~ Physics~ and~ Nuclear~ Engineering,~ P.O. ~ Box,
 MG-6,~077125 ~Magurele,~ Romania}
\address{$^b$ Horia~ Hulubei~ Foundation,~ P.O.~ Box,~ MG-12, 077125~ Magurele,~ Romania}

\ead{stoica@theory.nipne.ro}

\begin{abstract}
\\The rates of axion emission by nucleon-nucleon (NN) bremsstrahlung are
reconsidered by taking into account the NN short range correlations. The analytical formulas for 
the neutron-neutron (nn), proton-proton (pp) and neutron-proton (np) processes with the inclusion of the full momentum dependence of an one- and two- pion exchange nuclear potentials, in the non-degenerate limit, are explicitly given. We find that the two-pion exchange (short range) effects can give a significant contribution to the emission rates, and are temperature dependent. Other short range nuclear effects like effective nucleon mass, polarization effects and use of correlated wave functions, are discused as well. The trend of all these nuclear effects is to diminish the corresponding axion emission rates. Further, we estimate that the values of the emission rates calculated with the inclusion of all these effects can differ from the corresponding ones derived with constant nuclear matrix elements by a factor of $\sim 24$. This leads to an uncertainty factor of $\sim 4.9$ when extracting bounds of the axion parameters.

\end{abstract}

\vspace{2pc}

\submitto{\JPG}

\section{Introduction}

One of the most appealing solutions to the strong CP problem is the prediction of axions, pseudo Nambu-Goldstone bosons associated with the spontaneous breaking of the Peccei-Quinn symmetry \cite{[PQ77]}-\cite{[DFS81]}. 
Light axions might be emitted from all types of stars and might influence their evolution. 
In the lack of any experimental evidence, it is important to derive bounds on the axion properties, i.e. on their mass ($m_a$), on their couplings to matter ($g_{ai}$, a = axion and i = photon, electron, nucleon, etc.) and on the axion decay constant ($f_a$). These quantities are related by the equations:
$$f_a \simeq6 eV \frac{10^6 GeV}{m_a}\eqno(1)$$
$$g_{ai} \sim \frac{m_i N_q}{f_a}\eqno(2)$$

\noindent
where $N_q$ is the number of quark flavours.
These bounds can be obtained either from direct experiments or from astrophysical and cosmological considerations. For example, from the duration of the neutrino pulse from SN1987A one finds $m_a > 3 \times 10^{-3}$ eV \cite{[ERK97]}, while the requirement that axions do not overclose the universe leads to $m_a \ge 10^{-6}$ \cite{[SIK05]}. Bounds on $f_a$ ($m_a$) can also be extracted from the axion emission decay rates from newly born neutron stars during their cooling. The temperature and density in the core of such stars, just after their formation, can be T $\sim 25-80$ MeV and $\rho \simeq (6-8) \rho_0$, where {$\rho_0 \sim 2.5 \times 10^{14} g/cm^3$} is the ordinary nuclear matter density. In these extreme physical conditions only axions can excape from the core, and the dominant process for the axion emission is the NN bremsstrahlung: $NN \rightarrow \ NNa$ (N = neutron or proton). The decay rates of axion emission from these processes were previously calculated by different authors. Using an one pion exchange potential (OPEP) as NN interaction, Iwamoto calculated these emission rates for the degenerate 
(D) limit \cite {[IW8401]}. Using the nuclear matrix elements (NMEs) computed by Iwamoto, Turner calculated the emission rates for the non-degenerate (ND) limit \cite{[T88]} as well, while Brinkmann and Turner have performed  these calculations for arbitrary degeneracy using a constant term for the momentum dependent NMEs \cite{[BT88]}. This constant term was obtained by neglecting the pion mass as compared with the nucleon exchange momenta, in the expression of the spin summed squared matrix elements (the high momentum limit) \cite{[BT88]}. They concluded that the ND regime is a better approximation for the axion emissivities than the D regime, in the physical conditions characteristic for a newly born neutron star. The main difficulty in these previous calculations was the numerical evaluation of the multidimensional integrals appearing in the emission rates, when the full dependence of the NMEs on the nucleon momenta is included. In our previous work \cite{[SP09]} we developed a method of computing the axion emission rates by NN bremsstrahlung, with the inclusion of the full momentum contribution from a nuclear OPEP. We considered all the nn, pp and np bremsstrahlung processes in both the D and ND regimes, and found that the finite-momentum corrections to the emissivities are quantitatively significant for the ND regime and temperature-dependent. However, these calculations have neglected the influence of the dense nuclear medium. Ericson and Mathiot \cite{[EM89]} have investigated the modifications on the axion emission rates induced by short-range nuclear effects like the exchange of two pions between the nucleons (which can be simulated by the exchange of a (heavier) $\rho$-meson between nucleons), by many-body polarization effects and by the consideration of a nucleon effective mass. They found that under the physical conditions relevant for the axion emission from supernova SN1987A, all these effects can give significant corrections as compared to the simple estimations in terms of the OPEP, and obtained a much smaller lower bound for the axion decay constant. 
In this paper we try to improve the calculation of the axion emission rates by including the corrections induced by $\rho$-meson exchange between nucleons, which mimick the two-pion exchange effects. For that, we use our method developed in \cite{[SP09]}, which allows the inclusion in the numerical computation of the multidimensional integrals, of the full dependence on momenta of the nuclear potetial. Further, we estimate the uncertainty induced by the short-range nuclear effects when deriving of bounds of the axion parameters.

The paper is organized as follows: in section 2 we describe our method of calculation and obtain the theoretical formulae for the axion emission rates. In section 3 we present our results and comment on them, while the last section is devoted to the conclusions.

\section{Method of calculations}

The axion emission rate by NN bremsstrahlung is given by Fermi's
Golden Rule formula (see for instance \cite{[BT88]})

$$ \epsilon_{aNN}=(2\pi)^4 {\int \left[\Pi_1^4 \frac{d^3{\bf p}_i}{(2\pi)^3 2 E_i}\right]\frac{ d^3{\bf p}_{a}}{(2 \pi)^3 2E_a} E_a\times\left(S\times\Sigma{\vert{ M}\vert}^2 \right)\delta^4(P)\it F(f)} \eqno(3)$$

\noindent where,  ${\bf p}_i$ and $E_i$ (i=1,4) are the nucleon momenta and
energies, while ${\bf p}_a$ and $E_a$ are the corresponding axion
quantities; {\sl S} is a symmetry factor taking into account the
identity of the particles (S=1/4 for $nn$ and $pp$ channels and S=1
for the $np$ channel). 

${\it F(f)} = {\it f_1 f_2 (1-f_3) (1-f_4)}$ is
the product of Fermi-Dirac distribution functions of the initial
(1,2) and final (3,4) nucleons. In the non-relativistic limit $E_i \sim m + \frac{{\bf
p}_i^2}{2m}$ and using the non-dimensional quantities \cite{[BT88]}
$y_i = \hat{\mu_i}/T$ ($\mu_i$ are the chemical potentials of
the nucleons and T the temperature) and $u_i = {\bf p}_i^2 / 2mT$, the  expressions
of the Fermi-Dirac functions read ${\it f_i} = \left(\exp^{u_i -
y_i} + 1\right)^{-1}$. The degenerate (D) limit satisfies $y
>> 1$, while in the non-degenerate (ND) limit $y << - 1$. For the spin summed squared matrix elements (NMEs) we use  the
following expressions \cite{[YAK00]}:

$${ S}\times \Sigma {\vert {M}\vert^2} = { S}\times \frac{256}{3} g_{aN}^2 m^2 \left(\frac{f}{m_{\pi}}\right)^4 M_{NN} \eqno(4)$$

\noindent where $g_{aN}$ is the axion-nucleon coupling constant (we assume the same constant for the axion coupling to neutrons and protons). For the $nn$ and $pp$ the momentum-dependent factors $M_{NN}$ read \cite{[YAK00]}:

$$M_{nn}= \left(\frac{{\bf |k|}^2}{{\bf |k|}^2+m_{\pi}^2}\right)^2 +
\left(\frac{{\bf |l|}^2}{{\bf |l|}^2+m_{\pi}^2}\right)^2+\frac{(1-\beta){\bf |k|}^2 {\bf |l|}^2}{({\bf |k|}^2 +m_{\pi}^2)({\bf |l|}^2 + m_{\pi}^2)} \eqno(5)$$

\noindent while for the $np$ process
$$ M_{np} =\left(\frac{{\bf |k|}^2}{{\bf |k|}^2+m_{\pi}^2}\right)^2 + 2\left(
\frac{{\bf |l|}^2}{{\bf |l|}^2 + m_{\pi}^2}\right)^2-2(1-\beta)\frac{{\bf |k|}^2 {\bf |l|}^2}{({\bf |k|}^2+ m_{\pi}^2)({\bf|l|}^2 + m_{\pi}^2)} \eqno(6)$$

\noindent where ${\bf k} = {\bf p}_1 - {\bf p}_3 $ and ${\bf l} = {\bf
p}_1 - {\bf p}_4 $ are the nucleon direct and exchange transfer
momenta, respectively. The last (exchange) terms in the above
expressions arise from the interference of two different reaction
amplitudes. They contain contributions from the scalar product
$({\bf k\cdot l})^2$, which has been estimated (\cite{[BT88]}, \cite{[YAK00]}) by replacing
it with its phase-space averaged value, denoted by $\beta$. From kinematical constraints $\beta$ =0 in the D
regime, while it is $\sim 1.0845$ in the ND regime \cite{[RS95]}. Since the most important contribution to the emissivity comes from the ND limit, we perform our calculations only for this regime. 
For calculation, it is worth to estimate the contributions of the 
momentum exchange terms as compared with the pion mass in the physical conditions of axion emission. 
A typical nucleon momentum is $(3m_NT)^{1/2}$. At temperatures T $\sim$ few MeV the momentum exchange terms are only a few times larger than the pion mass term, and thus the pion mass cannot be neglected in the denominator of the NMEs expressions ((5) and (6)). Also, at small distances, below 2 fm, one expects that two-pion exchange effects become important. Their influence on the bremsstrahlung processes can be estimated by the exchange of a $\rho$ meson between the nucleons when the effective mass of this particle is taken to be $m^{*}_{\rho} \approx 600 $ MeV \cite{[HR98]}. When these effects are included, a typical term in the NMEs expressions is modified as follows:

$$\left(\frac{{|\bf k}|^2}{{|\bf k}|^2 + m_{\pi}^2}\right)^2 \rightarrow  \left(\frac{{|\bf k}|^2}{{|\bf k}|^2 
+ m_{\pi}^2} - C_{\rho} \frac{{|\bf k}|^2}{{|\bf k}|^2 + m_{\rho}^2}\right)^2 \eqno(7) $$ 

with $C_{\rho} = 1.67$ \cite{[HR98]}. 

For practical calculations we follow the procedure of Brinkmann and Turner \cite{[BT88]} to
derive the ND limit, by performing the transformation to the
center-of-mass system:

$$ {\bf p}_+ = \frac{{\bf p}_1 + {\bf p}_2}{2};~~{\bf p}_- = \frac{{\bf p}_1 -
{\bf p}_2}{2};$$ $$~~{\bf p}_{3c} = {\bf p}_3 + {\bf p}_+;~~{\bf p}_{4c} =
{\bf p}_4 + {\bf p}_+ $$
$$\Longrightarrow
{\bf p}_1={\bf p}_+ +{\bf p}_-;~~{\bf p}_2={\bf p}_+-{\bf p}_-;~~
{\bf p}_3={\bf p}_+ +{\bf p}_{3c};$$ $$~~{\bf p}_4= {\bf p}_++{\bf
p}_{4c}\Rightarrow {\bf p}_{4c} =- {\bf p}_{3c} \eqno(8) $$

and by defining the dimensionless quantities:

$$ u_i = \frac{{\bf p}_i^2}{2mT} (i={1,4});~~u_{\pm}=\frac{{\bf p}_{\pm}^2}{2mT};
~~u_{3c}=\frac{{\bf p}_{3c}^2}{2mT}, \eqno(9)$$
$$ cos{\gamma_1} = \frac{{\bf p}_+ {\bf p}_-}{\vert {\bf p}_+\vert \vert
{\bf p}_-\vert};~~cos{\gamma_c} = \frac{{\bf p}_+ {\bf p}_{3c}}{\vert
{\bf p}_+ \vert \vert {\bf p}_{3c}\vert};~~ cos{\gamma} = \frac{{\bf
p}_- {\bf p}_{3c}}{\vert {\bf p}_-\vert \vert {\bf
p}_{3c}\vert};\eqno(10) $$

 Following our method described in refs. \cite{[SP09]}, \cite{[SPN04]}, we express the NMEs contributions as a sum of three terms: a
 constant term, corresponding to the high-momentum limit when the pion mass is neglected as compared with the momentum transfer terms
 in the NMEs formula, and two separated correction terms representing the one- and the two-pion ($\rho$) contributions, when the full
 momentum dependence of the NMEs is taken into account. This way, we can better compare our results with the ones coming from previous
 calculations, where only the high-momentum limit contribution is considered \cite{[BT88]}. We did this by expressing the NMEs in
 terms of the scalar quantities $a_s = {\bf k}^2 + {\bf l}^2 = 2d(x + z/2)$ and $a_p = {\bf k}^2 {\bf l}^2 = d^2[x(x+2)sin^2\gamma +
 z^2/4]$, with $d=2mT$, $x = 2u_{3c}$ and $z = \omega/T$. In this way, one can arrange that the integral over the nuclear part to be
 reduced in a quite accurate approximation (within an error of $\sim 2\%$) into an integral which is independent of angles, and the
 integration can then be performed for the general case of the full momentum dependence of the NMEs. 

Finally, we expressed these NMEs in the following compact forms:
$${S}\times \Sigma {\vert {M}\vert^2} = \frac{64 m^2 g_{ai}^2}{3}\left(\frac{f}{m_{\pi}}\right)^4 \left[(3-\beta)-|M_{nn}^{\pi}|^2 -|M_{nn}^{2\pi}|^2\right] \eqno(11) $$
\noindent  for the nn and pp processes.

The corresponding one- and two-pion exchange corrections have the following expressions: 
$$ |M_{nn}^{\pi}|^2 = m^2_{\pi} \frac{A_{nn}^{\pi} - B_{nn}^{\pi} C_{\phi}^2}{C - D C_{\phi}^2 - E C_{\phi}^4} \eqno(12) $$ 

$$ |M_{nn}^{2\pi}|^2 = m^2_{\pi}\frac{A_{nn}^{2\pi} + B_{nn}^{2\pi}C_{\phi} + C_{nn}^{2\pi}C_{\phi}^{2} + D_{nn}^{2\pi}C_{\phi}^{3}}{E_{nn}^{2\pi} + F_{nn}^{2\pi}C_{\phi} + G_{nn}^{2\pi}C_{\phi}^{2} + H_{nn}^{2\pi}C_{\phi}^{3}} \eqno(13)$$

A similar procedure for the $np$ process gives:

$$ {S}\times \Sigma {\vert {M}\vert^2} = \frac{256 m^2 g^2_{aN}}{3}\left(\frac{f}{m_{\pi}}\right)^4 [\left(1+2\beta\right) -|M_{np}^{\pi}|^2-|M_{np}^{2\pi}|^2] \eqno(14) $$ 
where:
$$ |M_{np}^\pi|^2 = m^2_{\pi} \frac{A_{np}^{\pi} - C_{np}^{\pi}C_{\phi}^2}
{C - D C_{\phi}^2 - E C_{\phi}^4} +C_{\phi} \frac{B_{np}^{\pi} - D_{np}^{\pi}C_{\phi}^2}
{C - D C_{\phi}^2 - E C_{\phi}^4}\eqno(15) $$
\noindent and 
$$ |M_{np}^{2\pi}|^2 = |M_{nn}^{2\pi}|^2 \eqno(16)$$

The coefficients appearing in the above formula are polynomials depending on $m$, $T$ and $m_\pi$ and on variables $u_-$ and $u_{3c}$. The full expressions of the coefficents $A_{ij}^{\pi}$, $B_{ij}^{\pi}$, $C_{ij}^{\pi}$, $D_{ij}^{\pi}$, $C$, $D$ and $E$ can be found in \cite{[SP09]}, while the expressions of the coefficients $A_{nn}^{2\pi}$, $B_{nn}^{2\pi}$, $C_{nn}^{2\pi}$, $D_{nn}^{2\pi}$, $E_{nn}^{2\pi}$, $F_{nn}^{2\pi}$, $G_{nn}^{2\pi}$ and $H_{nn}^{2\pi}$ are given in the Appendix A. We also used the notation $C_{\phi}=cos \gamma_1 cos
\gamma_c + sin \gamma_1 sin \gamma_c cos\phi$, with $\phi$ the angle between the vectors ${\bf p}_+$ and ${\bf p}_-$.

Further, we could express the axion emission rates in the ND limit in a form with a similar structure as NMEs:

$$ \epsilon^{ND}_{ann} = \epsilon^{ND}_{ann}(0)\left(1-
\frac{I^{\pi}_{nn}}{(3-\beta)I_0} - \frac{I^{2 \pi}_{nn}}{(3-\beta)I_0}\right) \eqno(17) $$

\noindent where
$$\epsilon^{ND}_{ann}(0)= 2.68 \times 10^{-4}g^2_{an} e^{2y} m^{2.5} T^{6.5}(f/m_{\pi})^4 \eqno(18)$$
is the expression calculated by Brinkmann and Turner \cite{[BT88]}. 
$I_0$, $I^{\pi}_{nn}$  are double integrals over $u_-$ and $u_{3c}$, and are easily to be computed, while $I^{2\pi}_{nn}$ is a five-dimensional integral which needs a separate treatment.
$$ I_0 = \int_0^{\infty} \int_0^{u_-}{\sqrt{(u_-u_{3c})}(u_{-}- u_{3c})^2 e^{-2u_-}du_-du_{3c}} \eqno(19) $$

 $$ I^{\pi}_{nn} = \frac{\pi m_{\pi}^2}{mT}\int_0^{\infty} \int_0^{u_-} \sqrt{(u_-u_{3c})}(u_- - u_{3c})^2 e^{-2u_-} $$
$$\times \left( \frac{(7-\beta) + 4(3-\beta)(u_- + u_{3c})}{(2u_- + 2u_{3c} + m_1)^2} \right)du_- du_{3c}
\eqno(20) $$

$$ I^{2\pi}_{nn} =\int_{0}^{\infty}\int_{0}^{u_{3c}}\int_{0}^{\pi}\int_{0}^{\pi}\int_{0}^{2\pi}\left(u_{-}u_{3c}\right)^{\frac{1}{2}}\left(u_{-}-u_{3c}\right)^{2}e^{-2u_{-}}$$ 
$$\times \left(\frac{A_{nn}^{2\pi} + B_{nn}^{2\pi}C_{\phi} + C_{nn}^{2\pi}C_{\phi}^{2} + D_{nn}^{2\pi}C_{\phi}^{3}}{E_{nn}^{2\pi} + F_{nn}^{2\pi}C_{\phi} + G_{nn}^{2\pi}C_{\phi}^{2} + H_{nn}^{2\pi}C_{\phi}^{3}}\right)du_{-}du_{3c}d\gamma_{c}d\gamma_{1}d\phi  \eqno(21) $$

For the np process we find a final expression analogous to Eq.(17):

$$ \epsilon^{ND}_{anp} = \epsilon^{ND}_{anp}(0)\left((1+ 2 \beta) -
\frac{I^{\pi}_{np}}{3I_0} - \frac{I^{2\pi}_{np}}{3I_0}\right) \eqno(22) $$

\noindent with:

$$\epsilon^{ND}_{anp}(0)= 10.72 \times
10^{-4}g^2_{an} e^{y_{1}+y_{2}} m^{2.5} T^{6.5}(f/m_{\pi})^4 \eqno(23)$$

The correction integrals $I^{\pi}_{np}$ and $I^{2\pi}_{np}$ have the following expressions:

$$ I^{\pi}_{np} = \frac{\pi m_{\pi}^2}{mT}\int_0^{\infty} \int_0^{u_-}
\sqrt{(u_-u_{3c})}(u_- - u_{3c})^2 e^{-2u_-}\times $$
$$ \left(\frac{4(1+ 2 \beta)(u_-+u_{3c}) + (7 + 2 \beta)m_1}{(2u_- + 2u_{3c} +
m_1)^2} \right)du_- du_{3c} \eqno(24) $$

\noindent

$$ I^{2\pi}_{np} = I^{2\pi}_{nn}\eqno(25) $$

\section{Results}

We performed the integrals of the Eqs. (19), (20) and (24) without dificulty. To integrate the 5th-dimensional integral (21) we  built a special numerical routine based on multidimensional Gaussian quadratures, which is fast and higly accurate.
In order to better understand the error of neglecting the contributions of the momentum dependence of the spin summed squared NMEs, we plotted in Figs.~1-4 the temperature dependence of the relative axion emissivities ($(\epsilon(0) - \epsilon)/\epsilon(0)$) for the ND regime: for the nn/pp processes (Figs.~1-2) and for the np process (Figs.~3-4). We considered two values for the $\rho$ meson mass: i) the usual ``free'' mass $m_\rho = 770$ MeV, and ii) an effective mass $m^{*}_\rho= 600$ MeV, recommended when the meson exists in a dense nuclear medium \cite{[HR98]}. The lower curves from figures represent the case when only one pion exchange (OPE) effects are taken into account in calculation, while the upper curves represent the axion emissivities when both the OPE and the two-pion exchange (TPE) effects are included. One observes that the correction due to the TPE effects is quite important especially at higher T, while the total correction due to the inclusion of both effects is larger at lower T. Indeed, at T $\sim $ 75K the relative correction to the $\epsilon(0)$ due to OPE effects is  $\sim$ 30 $\%$, while this correction becomes $\sim 57\%$ when TPE corrections are included as well (see Fig.~2). The influence of the TPE effects diminishes at lower temperatures but the contribution of both of them to the emission rates calculated in the high-momentum limit increases to $\sim 97\%$ at T $\sim $ 25 K. Also, one can see that the TPE corrections depend, as expected, on the effective $\rho$ meson mass, being larger for the $\rho$ meson mass modified by the dense nuclear medium ($m^{*}_\rho = 600$ MeV). Another feature that we found is that these corrections are temperature dependent.

Finally, one remarks that the consideration of the full momentum dependence of an OPEP and TPEP has the general tendence of diminishing the axion emission rates, and the total relative error in computing the axion emission rates is between $30\%$ at $T \sim 75K$ and $97\%$ at $T \sim 25K$. This means that the axion emission rates might be overestimated up to a factor of 2, if one neglects the nuclear effects due to the full dependence on momenta of the nuclear potetial which includes both OPE and TPE effects.

However, in a dense nuclear medium, other nuclear effects should also be taken into account in calculation, as some of them are more important than those described above. These effects induce additional uncertainties in calculating the axion emission rates.
For example, nucleon effective masses are important, since the emission rates are proportional to $m_N^{2.5}$. The calculation of effective nucleon masses in a nuclear medium was subject of many works (see for example \cite{[BKS03]} and the references therein). In this sense, at nuclear densities $\rho \simeq 3\rho_0$, the effective nucleon mass can be half of its rest value: $m_N^{eff} \sim 0.5 m_N$. Thus, the emission rates can be overestimated by a factor of $\sim 6$. If one takes into account both uncertainities (full momentum dependence of an OPEP and TPEP and effective nucleon mass) the emission rates can be overestimated by a factor of $\sim 12$.
At high densities, the short range effects need also the calculation of transition amplitudes with correlated wave functions, instead of plane waves (Born approximation). The modified wave functions read:
$$\phi^c(r) = \phi^0(r) f(r);~~ f(r) = \sqrt{1-j_0(q_cr)}\eqno(26)$$

\noindent where $j_0$ is the Bessel function and the momentum $q_c \sim 800$ MeV/c fixes the scale of correlations. However, the modification in the nuclear potential due to this correction is rather small because of cancellation between $\pi$ and $\rho$ exchange effects. So, this correction is not expected to induce sizeable uncertainties to emission rates and axion parameters.
Other sizeable effects come from polarization effects of the $\pi$ and $\rho$ propagators from $\Delta$-hole excitations. A reduction of emission rates of a factor $\sim 2$ is expected when both the correlation function and the polarization effects are taken into account \cite{[EM89]}.
Summing up all these effects, we estimate an overall reduction of the emission rates by a factor of $\sim 24$, as compared with the calculation when only the high-momentum limit of an OPEP is considered. Since the emission rate is proportional to the square of the axion coupling constant, the overestimation of the value of the emission rate induces an uncertainty of a factor of $\sim 4.9$ when deriving the axion parameters.

\begin{figure}[tbh]
\begin{center}
\includegraphics[height=0.38\textheight,width=0.47\textwidth]{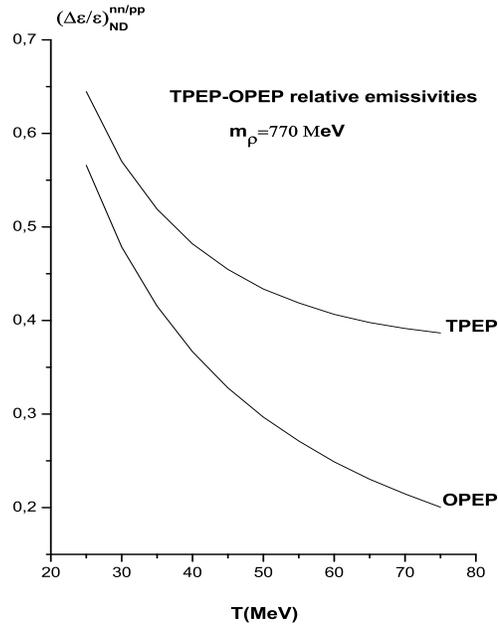}
 \end{center}
\caption{\label{fig:1}The axion relative emissivities from $nn,pp$ bremsstrahlung processes, in the ND limit, 
as a function of temperature $T$. The $\rho$ mass is taken $m_{\rho}=770 MeV$. The lower curve represents the case when only OPE effects are included in the calculation, while the upper curve represents the case when both OPE and TPE effects are included.}
\end{figure}

\begin{figure}[tbh]
\begin{center}
\includegraphics[height=0.38\textheight,width=0.47\textwidth]{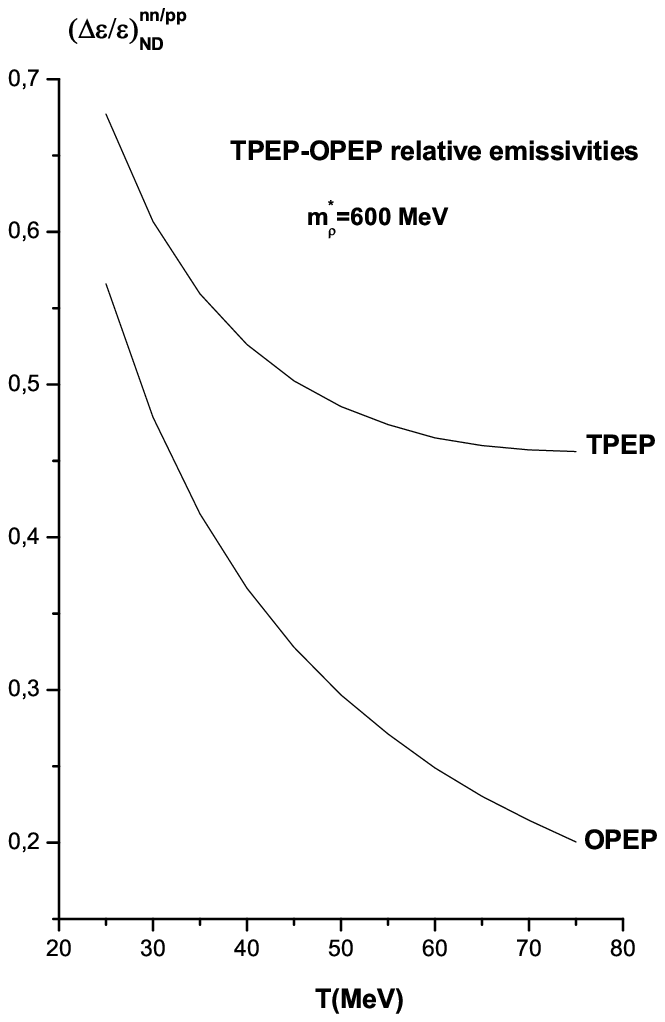}
 \end{center}
\caption{\label{fig:2}The same as in Figure 1, but for an effective mass $m^{*}_\rho= 600$ MeV }
\end{figure}

\begin{figure}[tbh]
\begin{center}
\includegraphics[height=0.38\textheight,width=0.47\textwidth]{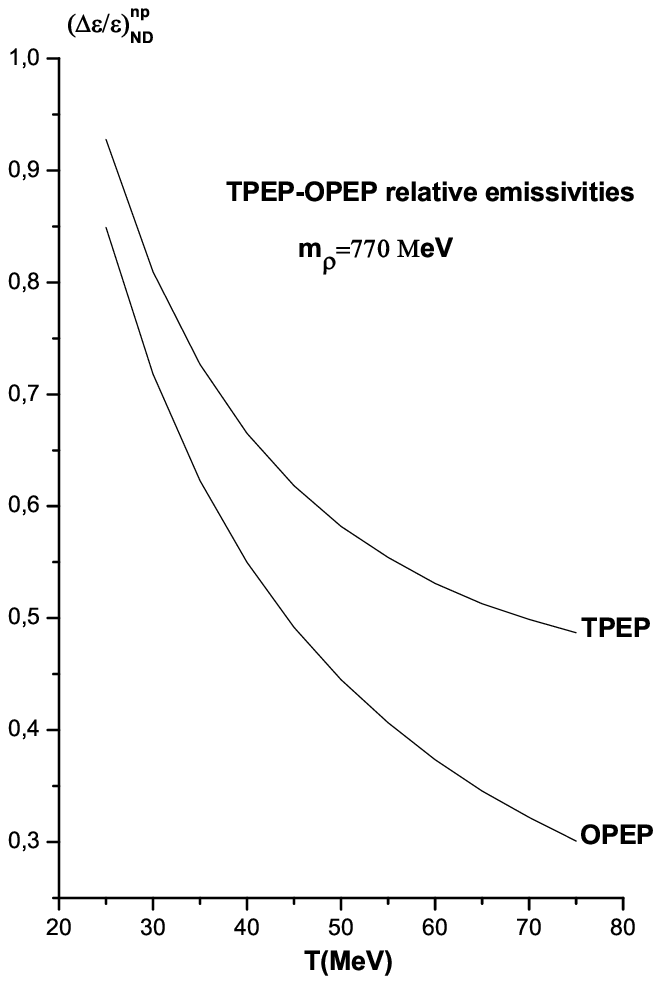}
 \end{center}
\caption{\label{fig:3}The axion relative emissivities from $np$
bremsstrahlung processes, in the ND limit, as a function of temperature $T$. The $\rho$ mass is taken $ m_{\rho}=770 MeV $. 
The lower curve represents the case when only OPE effects are included in the calculation, while the upper curve represents the case when both OPE and TPE effects are included.}
\end{figure}

\begin{figure}[tbh]
\begin{center}
\includegraphics[height=0.38\textheight,width=0.47\textwidth]{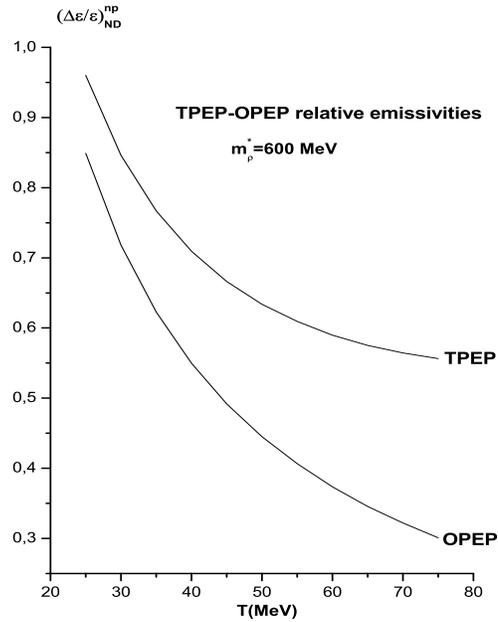}
 \end{center}
\caption{\label{fig:4}The same as in Figure 3, but for an effective mass $m^{*}_\rho= 600$ MeV }
\end{figure}

\section{Conclusions}

In this paper we calculated the axion emission rates by NN 
bremsstrahlung by taking into account the full dependence on momenta of a nuclear potential where both one- and two-pion exchange between nucleons were considered. The TPE were simulated by the exchange of a $\rho$ meson between the nucleons. We obtained analytical formulas for the nn, pp and np processes, in the ND regime, and perfomed the integrals over the momenta with a good numerical precision.  
We found substantial reductions of the emissions rates in comparison with the case when in calculation was considered only the high-momentum limit of the spin summed squared NMEs. The corrections due to TPE (short-range) effects are larger at higher temperatures, while at lower temperatures the influence on the emission rates is reduced. However, the inclusion of both OPE and TPE effects brings a significant contribution to the emission rates, being as well temperature dependent. At $T\sim 25$ K, it brings a correction of about $97\%$ to the values of axion emission rates calculated in the ``crude'' approximation of constant NMEs.
Also, one sees that the TPE corrections depend, as expected, on the effective $\rho$ meson mass, being larger for the $\rho$ meson mass modified by the dense nuclear medium ($m^{*}_\rho = 600$ MeV). 

We discussed as well other short range nuclear effects like effective nucleon mass, polarization effects and use of correlated wave functions. The trend of all these nuclear effects is to diminish the corresponding axion emission rates. Finally, we estimated the overall uncertainty in computing of the emission rates related to all these nuclear effects. We found that the values of the axion emission rates calculated with the inclusion of all these effects can differ from the coresponding ones calculated with constant NMEs, by a factor of $\sim 24$. This leads to an uncertainty by a factor of $\sim 4.9$ when extracting bounds for the axion parameters.

\section{Acknowledgements}

We aknowledge the support of the project IDEI-975/2008.

\section{Appendix A}

We present here a part of the coefficients entering in the correction matrix elements for the cases which we presented in the section 2.

$$ A_{nn}^{2\pi}=8C_{\rho}m^{3}T^{3}\left(u_{-}+u_{3c} \right)^{3}\left( C_{\rho}-2\right)+4C_{\rho}m^{2}T^{2}\left(u_{-}+u_{3c} \right)^{2}\left(m_{\pi}^2C_{\rho}-2m_{\rho}^2 \right) \eqno(A.1)$$

$$ B_{nn}^{2\pi}=-48C_{\rho}m^{3}T^{3}\left(u_{-}+u_{3c} \right)^{2}\left(u_{-}u_{3c}\right)^\frac{1}{2}\left( C_{\rho}-2\right)+16C_{\rho}m^{2}T^{2}$$ 
$$\times\left(u_{-}+u_{3c} \right)\left(u_{-}u_{3c}\right)^\frac{1}{2}\left(-C_{\rho}m_{\pi}^{2}+2m_{\rho}^2 \right) \eqno(A.2)$$

$$C_{nn}^{2\pi}=64C_{\rho}m^{3}T^{3}\left(u_{-}+u_{3c}\right)\left(u_{-}u_{3c}\right)\left(C_{\rho}-2\right)+$$
$$ +16C_{\rho}m^{2}T^{2}\left(u_{-}u_{3c}\right)\left(C_{\rho}m_{\pi}^{2}-2m_{\rho}^2\right)\eqno(A.3)$$

$$ D_{nn}^{2\pi}=-64C_{\rho}m^{3}T^{3}\left(u_{-}u_{3c}\right)^\frac{3}{2}\left( C_{\rho}-2\right)\eqno(A.4) $$

$$ E_{nn}^{2\pi}= 8 m^{3}T^{3}\left(u_{-}+u_{3c}\right)^{3}+4m^{2}T^{2}\left(u_{-}+u_{3c}\right)^{2}\left(m_{\pi}^2 + 2m_{\rho}^2 \right)+ $$ 
$$+2mT\left(u_{-}+u_{3c}\right)m_{\rho}^{2}\left(m_{\rho}^2+2m_{\pi}^2 \right)+m_{\pi}^2m_{\rho}^{4} 
\eqno(A.5)$$

$$F_{nn}^{2\pi}=32m^{3}T^{3}\left(u_{-}+u_{3c}\right)^{2}\left(u_{-}u_{3c}\right)^{\frac{1}{2}}+16m^{2}T^{2}\left(u_{-}+u_{3c}\right)^{2} \left(u_{-}u_{3c}\right)^{\frac{1}{2}}+$$
$$+16m^{2}T^{2}\left(u_{-}+u_{3c}\right)\left(u_{-}u_{3c}\right)^{\frac{1}{2}}\left(m_{\pi}^2+2m_{\rho}^2\right)+4mT\left(u_{-}u_{3c}\right)^{\frac{1}{2}}m_{\rho}^{2}\left(m_{\rho}^2+2m_{\pi}^2 \right)  \eqno(A.6)$$ 

$$G_{nn}^{2\pi}=96m^{3}T^{3}\left(u_{-}+u_{3c}\right)\left(u_{-}u_{3c}\right)+16m^{2}T^{2}\left(u_{-}u_{3c}\right)\left(m_{\pi}^2+2m_{\rho}^2 \right) \eqno(A.7)$$ 

$$ H_{nn}^{2\pi}=-64m^{3}T^{3}\left(u_{-}u_{3c}\right)^{\frac{3}{2}} \eqno(A.8) $$

\end{document}